# The tardigrade as an emerging model organism for systems neuroscience

## Graphical Abstract

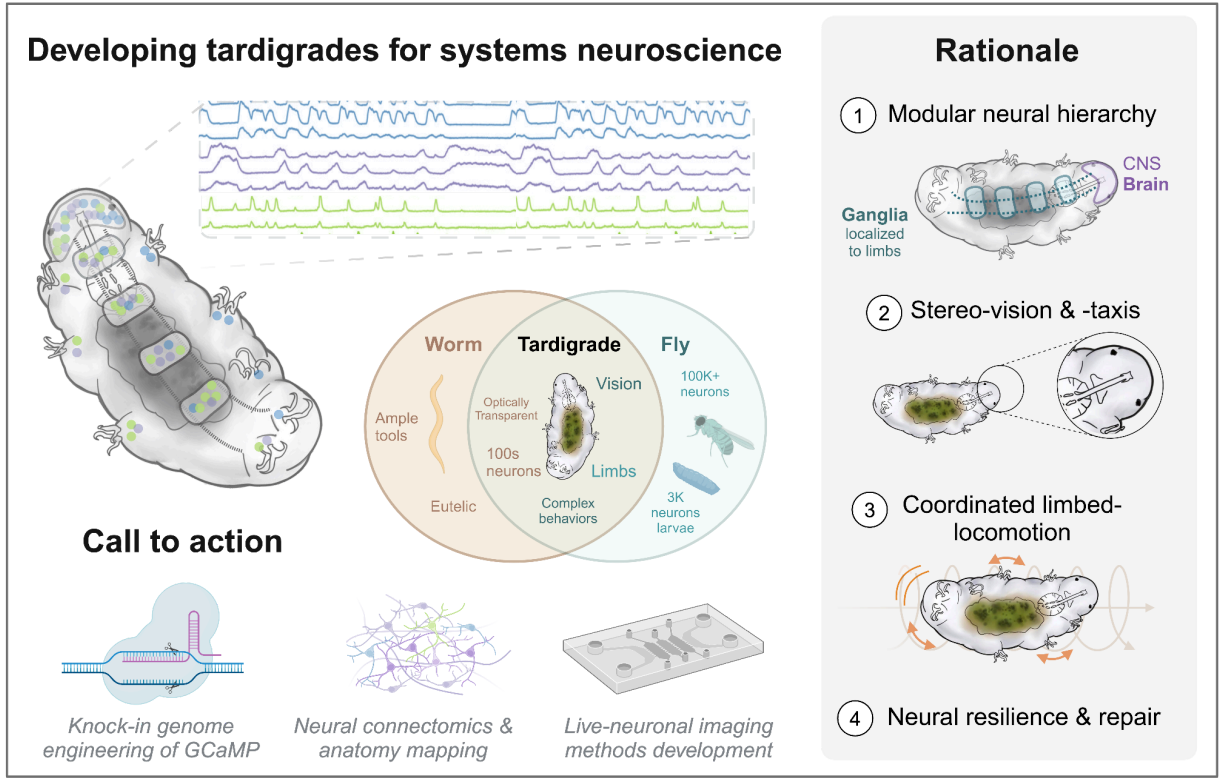

## Authors
Ana M. Lyons, Saul Kato

## Correspondence
ana.marie.lyons@gmail.com
saul.kato@ucsf.edu

## Highlights

- Tardigrades are emerging model organisms that exhibit complex behaviors with a nervous system of only several hundred neurons

- These optically transparent microscopic animals enable novel studies of limbed coordination, hierarchical control, stereo vision, and neural resilience

- Their modular nervous system bridges simple invertebrates and more complex systems

- Call to action: genetic tools and neuroanatomical mapping are needed

# The tardigrade as an emerging model organism for systems neuroscience

Ana M. Lyons[1], Saul Kato[1]

[1] Foundations of Cognition Lab, Department of Neurology, Weill Institute for Neurosciences, University of California, San Francisco, USA
Correspondence: ana.marie.lyons@gmail.com, saul.kato@ucsf.edu

**SUMMARY**

We present the case for developing the tardigrade (*Hypsibius exemplaris*) into a model organism for systems neuroscience. These microscopic, transparent animals (~300-500 μm) are among the smallest known to possess both limbs (eight) and eyes (two), with a nervous system of only a few hundred neurons organized into a multi-lobed brain, ventral nerve cord, and a series of ganglia along the body. Despite their neuroanatomical simplicity, tardigrades exhibit complex behaviors, including multi-limbed walking gaits, individual limb grasping, phototaxis, and transitions between active and dormant states. These behaviors position tardigrades as a uniquely powerful system for addressing certain fundamental questions in systems neuroscience, such as: How do nervous systems coordinate multi-limbed behaviors? How are top-down and bottom-up motor control systems integrated? How is stereovision-guided navigation implemented? What mechanisms underlie neural resilience and recovery during environmental stress? We review current knowledge of tardigrade neuroanatomy, behavior, and genomics, and we identify opportunities and challenges for leveraging their unique biology. We propose developing essential neuroscientific tools for tardigrades, including genetic engineering and live neuroimaging, alongside behavioral assays linking neural activity to outputs. Leveraging their evolutionary ties to *C. elegans* and *Drosophila melanogaster*, we can adapt existing toolkits to accelerate tardigrade research—providing a bridge between simpler invertebrate systems and more complex neural architectures.

**INTRODUCTION**

Systems neuroscience aims to understand how the structure of the nervous system produces network activity, and how this activity in turn produces behavior. Behavior may be thought of as the process of gathering multi-sensory input, integrating it with a dynamic internal state, and generating both reflexive as well as goal-directed, complex motor actions. Established model organisms have revolutionized this field, from the relative biological simplicity of *C. elegans*, which offers high fidelity single-neuron resolution imaging of whole animals, to the behavioral richness of *Drosophila*, zebrafish, and mice. Yet, a gap persists: can we comprehensively monitor, manipulate, and model neural activity at a systems level in organisms capable of computationally complex, goal-directed behaviors, but with simpler, more tractable, nervous systems? Here, we propose that tardigrades (*Hypsibius exemplaris*), also known as water bears, offer an unprecedented opportunity to fill this gap—standing poised to become rising stars in the pantheon of systems neuroscience model organisms.

Studying animal model organisms of varying brain complexity has been essential to the field of systems-level neuroscience since its dawn. There is a general tradeoff between the accessibility of neural systems for study and the cognitive sophistication of animal brains, with each model offering distinct advantages for addressing particular neuroscientific questions. Tardigrades uniquely display both anatomical simplicity and behavioral complexity, while nestled in among *Panarthropoda* in their own phylum (**Figure 1**A), offering a unique opportunity to understand how a compact nervous system orchestrates sophisticated behaviors (Gross et al., 2019; Schill, 2019). These microscopic animals (~200-1500 μm in length) have a nervous system of only a few hundred neurons, including a central brain, two simple eye-spots, a ventral nerve cord, and ganglia connected to four pairs of legs (**Figure 1**B, **C**) (Gross et al., 2019; Møbjerg et al., 2018). Despite this anatomical simplicity, tardigrades perform remarkably complex, goal-directed behaviors such as multi-limbed gait coordination, individual limb use, dormancy transitions, phototaxis, thermotaxis, and chemotaxis—all suggesting computationally and cognitively rich neural control (Nirody et al., 2021; Beasley 2021; Anderson et al., 2024; Hvidepil and Møbjerg 2023; Loeffelholz et al., 2024). Their compact nervous system and transparent body offers accessibility for live neural activity imaging and perturbation.

Given their biological diversity and successful colonization of virtually every biome, tardigrades have long been favored animals of study for taxonomists, ecologists, and evolutionary developmental biologists (Nelson et al, 2018). However, they have largely escaped the attention of neuroscientists. Establishing tardigrades as a robust model for systems neuroscience requires investment in adapting methodologies originally developed for other organisms. The anatomical and functional mapping of the tardigrade nervous system is in its infancy: the exact number of neurons in any tardigrade life-history stage remains unknown, neuronal identities and types have not been established, their connectome has yet to be reconstructed, and the mechanisms underlying the animal's fascinating neurobiology, such as neuronal resilience and repair, remain unexplored (see **Table 1** for current neuroanatomical knowledge). **Figure 2** highlights historical and modern visualizations of the tardigrade nervous system, showing how emerging techniques like *in situ* hybridization and immunohistochemistry are advancing our understanding of their neural architecture.

Historically, the development of new model organisms has required decades of painstaking biological study and methods development, resulting in a scarcity of experimental organisms. **Table 2** highlights the unique traits of tardigrades as emerging models, contrasting their advantages and current challenges with other established systems neuroscience models. Advances in genetic manipulation technologies such as CRISPR, neural activity indicators and effectors like GCaMP and channelrhodopsin, and high-resolution imaging tools—combined with the explosion of computational capability—have made it feasible to establish new model organisms in a fraction of that time. Moreover, tardigrades' sensory modalities and the potential for innovative behavioral assays, as outlined in **Table 3**, position them as a promising system for studying sensory integration and adaptive responses.

By adapting methodologies originally developed for other organisms, tardigrades can tackle fundamental systems neuroscience questions not easily approached in other model organisms. We call upon the community to embrace this opportunity, recognizing tardigrades as a vital addition to the expanding toolkit of neuroscience research.

## I. SCIENTIFIC RATIONALE FOR TARDIGRADES FOR SYSTEMS NEUROSCIENCE: AREAS OF INQUIRY

We present several systems neuroscience questions that are particularly primed for study in the tardigrade (**Figure 3**).

### How do nervous systems coordinate multi-limbed behaviors?

Tardigrades possess a modular nervous system that combines a central brain with four segmental ganglia, each anatomically associated with a pair of legs. This organization allows for the generation and coordination of complex locomotion, including multi-limbed gaits and transitions between behavioral states. Unlike the undulatory crawling motion of *C. elegans*, tardigrades exhibit limb-based gaits, reminiscent of arthropods and mammals. While fruit fly larvae (*D. melanogaster*) crawl with articulated body segments, they lack sophisticated limbs, gaits, or grasping behaviors, and they have a nervous system of over 5000 neurons, an order of magnitude larger than tardigrades (Winding et al., 2023). In the tardigrade, we may ask: how are complex, adaptive gaits implemented by a neuromuscular system? Are there pattern generator subsystems, and if so, how do they support differing gait patterns? How are gaits modulated to serve goal-directed behaviors? Investigation in tardigrades could offer unprecedented insights into how modular circuits generate complex, adaptive movement patterns.

### How do top-down and bottom-up motor control systems interact?

The tardigrade nervous system presents an exceptional opportunity to dissect the interplay between localized and centralized control of behavior. Each limb of the tardigrade participates in coordinated walking gaits in the service of goal-oriented taxis, but also shows its own independent swinging and grasping movements. Each ganglion may function as a semi-autonomous computational unit, capable of processing sensory inputs and generating motor programs at a single limb level, while also being recruitable into coordinated gait programs. How the integration of these movement modes produces cohesive whole-body behavior remains unknown. How do sensory signals from individual limbs or sensory structures influence ganglia-level computations, and how are these signals integrated with top-down signals? How do tardigrades break out of rhythmic gaits for individual limb movement? What role does feedback from the brain play in fine-tuning the outputs of individual ganglia? Tardigrades, with their modular nervous system and anatomically distinct ganglia, offer an unparalleled model for probing the hierarchical integration of neural circuits and the dynamic balance between local autonomy and global control.

### How does stereo light sensing drive taxis behavior?

Tardigrades have remarkable sensory capabilities that invite investigation from a systems neuroscience perspective. Their rudimentary eyespots, consisting of clusters of photoreceptive cells, guide behaviors such as phototaxis. Unlike *C. elegans*, which lacks apparent vision-specific anatomical or neural structures, tardigrades possess two laterally separated eyespots, suggesting the capability for stereo-vision. How do tardigrades process stereo visual inputs to navigate their environments? Are eyespots evolutionary precursors to more sophisticated visual systems and is there conservation of neural implementations of directional navigation? Additionally, tardigrades respond to multiple sensory modalities, including chemical and thermal stimuli, making them an excellent system for exploring how diverse sensory inputs are integrated to drive complex behaviors.

### What can tardigrades reveal about neural resilience and dormancy?

Tardigrades are renowned for their ability to survive extreme environmental conditions, including desiccation, freezing, and radiation. The mechanisms underlying dormancy initiation, maintenance, and recovery remain poorly understood. How does the tardigrade nervous system respond to environmental stressors to initiate dormancy? What adaptations enable neurons to maintain function during metabolic shutdown? How do tardigrades repair neural damage caused by environmental stress, and what molecular mechanisms confer their remarkable resilience? What and how are memories or conditioned neural states maintained through periods of low or absent metabolic activity? Answering these questions could have profound implications for biomedicine, particularly in the context of neuroprotection and repair. The tardigrade's ability to endure and recover from extreme conditions offers a unique opportunity for exploring the limits of neural plasticity and resilience.

## II. CALL TO ACTION: BUILDING THE TOOLKIT FOR TARDIGRADE NEUROSCIENCE

For tardigrades to fulfill their potential as a model system in systems neuroscience, the development of specific tools and methodologies is essential. This effort requires a coordinated approach to overcome existing bottlenecks and address gaps in molecular biology, neuroanatomy, live imaging, and behavioral analysis. By investing in these areas, we can establish tardigrades as a transformative model for exploring neural principles across scales, from single neurons to whole systems:

### Molecular biology, genomics, and genetic engineering tools

The development of robust genetic tools is a critical frontier for tardigrade research. While approaches from *C. elegans*, *Drosophila*, and other models provide launching points, tardigrades present unique obstacles, particularly in genetic cargo delivery, promoter optimization, and transgene stability. Resolving these bottlenecks will require innovative adaptations tailored to tardigrades' unique physiology.

*Transgenics* The ability to express exogenous genetic constructs is a triumph of diverse utility in molecular biology, enabling precise dissection of gene function and cellular dynamics. In particular, neuronal expression of genetically encoded activity sensors, such as GCaMP, allows real-time visualization of neural activity. In *C. elegans*, transgenics are most commonly achieved through the generation of extrachromosomal arrays, which are formed by microinjecting DNA into the gonad. These arrays are maintained episomally and allow for rapid production of transgenic lines (Mello et al., 1991). For stable integration, methods such as Mos1-mediated single-copy insertion (MosSCI) and CRISPR-Cas9 are employed, offering precise, single-copy transgene insertion into the genome (Frøkjaer-Jensen et al., 2008; Dickinson et al., 2015)

The most pressing challenge is genetic cargo delivery—how to efficiently introduce transgenes into tardigrade cells (**Figure 4**). Electroporation protocols, which are routine in *C. elegans* and *Danio rerio* (zebrafish) embryos, remain underdeveloped for tardigrades and require optimization for osmotic compatibility, cuticle integrity, and life-stage variability (Khodakova et al., 2021; Buono and Linser, 1998; Zhang et al., 2020). Viral infection methods and chemical transfection reagents may provide alternative approaches, but their effectiveness in tardigrades has not been tested (Li et al, 2013; Aronovich et al., 2011).

Developmental stages amenable for genetic modification—akin to blastoderm injections in *Drosophila* or early embryo editing in *C. elegans*—is another unknown (Guidetti, 2024). Given the limitations of embryonic manipulation, delivery strategies for juveniles or adults, such as direct parental CRISPR (DIPA-CRISPR) could be a promising alternative (Kondo et al., 2024).

The choice of genetic cargo and necessary cofactors is an open question. Transposon systems like PiggyBac and Sleeping Beauty, have been successful in many other invertebrate systems, while the new entrant CRISPR has shown efficacy across the tree of life. The particular choice of CRISPR system itself would require further evaluation. While Cas9 has been used in initial studies, its efficiency in tardigrades is low. Cas12a (Cpf1), with its thermal stability and staggered DNA cuts that improve homology-directed repair, could be better suited for tardigrades given their resilience to environmental stressors (Kleinstiver et al., 2019).

Promoter optimization is another need. In *C. elegans* and *Drosophila*, pan-neuronal promoters like *rab-3*, *synaptobrevin*, and *elav* have enabled precise tissue-specific expression of fluorophores and activity indicators (Hobert Lab; Bhattacharya et al., 2002). In tardigrades, the TardiVec system has demonstrated transient expression using 1-kb upstream regions of ubiquitously expressed genes like actin and tubulin (Tanaka et al., 2023). However, it is unclear whether this promoter length is optimal, or if regulatory sequences in tardigrades are uniquely compact. Systematic screening of additional promoters, particularly for tissue-specific activity, is needed. Identifying a ubiquitous 3'UTR that stabilizes transcripts and enhances transgene expression across tissues and developmental stages remains equally unknown.

Reliable co-injection or co-transfection markers are also essential for tracking genetic cargo delivery and expression. In *C. elegans*, ubiquitous markers like GFP driven by *myo-2* or *unc-54* promoters validate successful microinjection, while non-overlapping fluorophores distinguish experimental targets (Okkema et al., 1993; Moerman et al., 1982). Tardigrades' natural autofluorescence in the gut and cuticle presents a non-trivial challenge, particularly in older animals or specific life stages (Bartels et al., 2023). Identifying fluorophores in the far-red or near-infrared spectrum that bypass autofluorescence will be crucial for improving signal detection.

Identifying benchmark genes whose knockout produce a clear and scorable phenotype, analogous to *unc-54* or *elav* in other animals, would enable systematic testing of gene editing tools and delivery methods. In *C. elegans*, automated worm sorters and imaging systems have enabled rapid behavioral and phenotypic screening across many thousands of animals (O'Reilly et al., 2016). Similar platforms tailored to tardigrades—integrating microfluidics with high-resolution imaging, alongside high-throughput methods to mutagenized animals—could streamline transgene screening and phenotyping.

*Mutagenesis* Mutagenesis has been pivotal in propelling research in *C. elegans* and *Drosophila*, allowing for functional genetic screens and the elucidation of developmental, neural, and behavioral pathways. In *C. elegans*, chemical mutagens like EMS (ethyl methanesulfonate) and UV radiation have generated libraries of mutants that revolutionized our understanding of neural connectivity and behavior (Brenner, 1974). Similarly, in *Drosophila*, random insertional mutagenesis with transposons has been instrumental in gene discovery and functional analysis (Rubin and Spradling, 1982). These techniques have enabled studies of genetic pathways underlying diseases like Parkinson's, Alzheimer's, and cancer.

For tardigrades, similar mutagenesis strategies could be adapted using radiation (e.g., gamma or UV), chemical mutagens, or targeted genome editing technologies such as CRISPR. Radiation-based mutagenesis may offer insights into tardigrades' exceptional DNA repair mechanisms, which could be exploited to study human diseases related to genome stability, such as cancer or neurodegenerative disorders. Chemical mutagens could uncover genes involved in stress tolerance, locomotion, or neural resilience, enabling targeted follow-up studies.

*Pitfalls* Achieving transgenesis in new model organisms is often fraught with species-specific impediments and unknowns, requiring customization of methods found through trial and error. For example, small RNA-mediated silencing during early development could interfere with genome editing or transgene expression. In other systems such as *C. elegans*, small RNAs regulate early gene expression and may compete with exogenous constructs (Ambros et al., 2003). MicroRNA sequencing across tardigrade developmental stages would help identify potential silencing pathways and inform strategies to circumvent potential interference.

In summary, **genetic tool development** needs include:

- *An effective **genetic cargo delivery** method for genetic cargo into embryos or mothers*

- *An effective DNA modification method such as **transposon** (PiggyBac) or **CRISPR** systems (Cas9, Cas12a, or others)*

- *Reliable **mutagenesis** methods*

- ***Promoter optimization** for reliable, tissue-specific expression. What **benchmarking genes** should be used?*

### Neuroanatomy: mapping the tardigrade nervous system

Neuroanatomical mapping at multiple levels, from synaptic connectivity to gross anatomy, is of critical value for systems neuroscience. A tardigrade connectome would be a key resource. In *C. elegans*, the production of electron microscopy (EM)-based connectomes revealed the complete synaptic architecture of its 302 neurons, directly linking anatomical circuits to specific behaviors (White et al., 1986). Similarly, in *Drosophila*, electron microscopy has resolved synaptic connections at nanometer resolution, generating detailed maps of neuron cell types and neural circuits across developmental stages and brain regions (Dorkenwald et al., 2024, Schlegel et al., 2024, Zheng et al., 2018).

The construction of a statistical atlas of neural positions is a related goal for mapping tardigrade neuroanatomy, similar to efforts in *C. elegans* (Sprague et al., 2024). In *Drosophila*, automated segmentation of high-resolution confocal and EM datasets has produced standardized brain atlases that integrate anatomical data with functional and behavioral studies (Chiang et al., 2011). Advances in machine learning and AI-driven segmentation tools have accelerated the analysis of large imaging datasets in other models and could be applied to the tardigrade.

Expansion microscopy, which physically expands tissues to enhance imaging resolution, could be particularly powerful in tardigrades for visualizing their small neurons and synapses while preserving tissue architecture. Combining ExM with fluorescent protein labeling, immunostaining, and high-resolution light microscopy could enable synapse-resolution mapping of neural connections across the brain, ventral nerve cord, and segmental ganglia, as in *C. elegans* (Yu et al., 2020).

A pressing question in tardigrade neuroanatomy is the relationship between neural structures and sensory appendages, such as eyespots, claws, and cephalic mechanosensory structures. In *C. elegans* and *Drosophila*, sensory neurons have been mapped with high precision, linking specific sensory inputs to synaptic targets and downstream motor behaviors (Chalfie et al., 1994; Lazar et al., 2021, Schafer, 2005). In tardigrades, however, the molecular identity and connectivity of sensory neurons remain poorly understood. Tools such as single-cell RNA sequencing can be applied to identify sensory neuron subtypes and the genes they express.

A salient unknown in tardigrades is the question of **neuro-eutely**—whether the total number of neurons is precisely specified in development, as it is in *C. elegans* (Sulston and Horvitz, 1977). Resolving this question requires cell lineage analyses and developmental time-course studies using immunolabeling, single-cell transcriptomics, and live imaging, beyond existing preliminary work (Quiroga-Artigas and Moriel-Carretero, 2024). Additionally, these tools can help identify the differentiation pathways of neuronal and non-neuronal cells—establishing the complete cell lineage tree of the worm was a major catalyst for *C. elegans* neuroscience and Nobel win (Sulston et al., 1983).

> In summary, **tardigrade neuroanatomy** needs include:
>
> - ***Synaptic resolution** structural imaging tools for of the brain, ventral nerve cord, and ganglia using EM, light microscopy, possibly facilitated by expansion microscopy*
>
> - *Construction of a **3D neuron statistical atlas** by integrating high-resolution imaging and automated segmentation*
>
> - *Identification of **molecular** and **functional identities** in sensory structures and motor circuits using **single-cell RNA sequencing** and/or **proteomics***
>
> - *Development of a **connectome** to map neural architecture. Resolve the question of **neuro-eutely** across life stages or in response to stress.*

### Neural activity imaging and behavior

The ability to observe neural activity in real time is a cornerstone of systems neuroscience, and the integration of high-resolution neural imaging with robust behavioral assays is of particular value. By leveraging tools and approaches developed in *C. elegans* and *Drosophila*, such as volumetric microscopy, microfluidic systems, and computational methods for extracting neural activity time series, we can address these challenges and unlock the potential of tardigrades for real-time neural imaging.

In *C. elegans*, calcium imaging using GCaMP has enabled the monitoring of neural activity with single-cell resolution (Schrödel et al., 2013; Kato et al., 2015) (**Figure 5**A). This approach can be readily adapted for tardigrades, which are optically transparent and small enough to facilitate high-resolution imaging of neural dynamics (**Figure 5**B). As with *C. elegans*, to support such imaging, microfluidic chambers optimized for tardigrade immobilization will be required to enable immobilization for high quality imaging while allowing controlled sensory stimulation (Gray et al., 2005; Pokala & Flavell, 2022).

Free-moving imaging systems, which integrate live animal tracking with neuroimaging, are also essential. Tools such as worm trackers and behavioral analysis pipelines in *C. elegans* enable precise quantification of locomotion and responses to stimuli (Pokala & Flavell, 2022). Achieving real-time imaging of neural activity in freely moving tardigrades would allow researchers to study the neural basis of behaviors like phototaxis, chemotaxis, and multi-limbed locomotion.

Nuclear localization of calcium indicators has been a highly successful strategy in many organisms to enable neuron-resolution whole brain imaging, first established in *C. elegans*, by facilitating image segmentation of neurons (Schrödel et al., 2013). Building on this strategy, tools like NeuroPAL heterogeneous color labeling and automated segmentation pipelines have enabled the accurate identification and tracking of neurons across time in immobilized and freely moving animals (Yemini et al., 2021; Sprague et al., 2024). By analyzing brain-wide activity patterns, distributed neural networks in *C. elegans* were found to coordinate transitions between behavioral states, such as locomotion and quiescence (Kato et al., 2015; Nichols et al., 2017). Similar approaches can be applied in tardigrades to yield insights in the neural production of behavior.

Optogenetic manipulation will be crucial for establishing causal roles of neurons. In *C. elegans*, optogenetic tools like channelrhodopsins have been used to understand sensory-guided decision making (Dunn et al., 2024). Similarly, *Drosophila* studies have employed optogenetics to understand how sensory neurons in the antennae and compound eyes drive phototaxis and grooming behaviors (Zhang et al., 2020; Riemensperger et al, 2016). Adapting optogenetics for tardigrades will require advances in transgene delivery and expression, particularly for fluorophore-tagged opsins under tissue-specific promoters. Such tools could clarify how sensory structures like eyespots and mechanoreceptors contribute to specific adaptive behaviors.

Reliable, quantitative behavioral assays will be essential for connecting neural activity to normative function. In *C. elegans*, automated tracking systems have been used to quantify behavior, including responses to sensory stimuli, such as escape behaviors or movement toward chemical gradients (Barlow et al., 2022, Albrecht and Bargmann, 2011). Developing similar high-throughput behavioral platforms for tardigrades—integrating controlled sensory stimulation with automated tracking—will allow for systematic studies of sensory-driven behaviors, including phototaxis, chemotaxis, and responses to mechanical cues. These assays will provide a scalable means of testing sensory function, neural integration, and adaptive responses under different experimental conditions.

In summary, needs for **live neural imaging** include:

- ***Transgenics*** *for expressing genetically encoded activity indicators and optogenetic effectors*
- *Development of **microfluidic chambers** for movement constraint and controlled sensory stimulation*
- ***Volumetric imaging of neural activity*** *in immobilized and freely moving animals*
- *Computational pipeline for **neuronal segmentation and tracking***
- ***Quantitative behavioral assays***

### Community resources

Establishing community resources will be essential for scaling tardigrade research and ensuring accessibility, efficiency, reproducibility, and innovation. We aim to model successful research communities like those of *C. elegans* and *Drosophila*.

*Bioinformatics and reagent sharing* A central repository for wild-type and transgenic tardigrade strains is a foundational need. In *C. elegans*, resources such as WormBase and the Caenorhabditis Genetics Center (CGC) and have enabled researchers to aggregate biological data and distribute actual frozen worm strains, facilitating progress and experimental reproducibility across labs (Harris et al., 2020) and long-term storage for biodiversity preservation. Similarly, for *Drosophila*, the FlyBase and Bloomington *Drosophila* Stock Center aggregate gene data and distribute mutant and transgenic lines (Ozturk-Colask, et al. 2024). Tardigrade genomic data should be integrated into established platforms such as CRISPR guide RNA design tools (e.g., CHOPCHOP, Benchling, or CRISPOR) (Labun et al., 2019; Montague et al., 2014; Concordet et al., 2018). In addition, sequenced genomes should be publicly available with genome browsers, similar to WormBase or FlyBase. A crowd-sourced repository of tested promoters would further catalyze the development of genetic tools. In *C. elegans* and *Drosophila*, validated tissue-specific promoters are widely shared through databases and community resources.

*Protocol sharing* Detailed protocols for molecular biology, culturing and maintaining tardigrades should be aggregated. Protocols for wild-type strain propagation, handling during imaging experiments, and stress or behavioral assays should be collaboratively developed, curated, and shared in open-access formats. Similar efforts in the *C. elegans* community have resulted in accessible, peer-reviewed protocols (e.g., through the WormBook and Protocols.io) that ensure consistent practices across laboratories. The foundational role of community resources like The Worm Breeder's Gazette was instrumental in the early days of worm research, fostering collaboration and innovation among researchers. Recently relaunched, it continues to support the dissemination of protocols, insights, and updates within the *C. elegans* community, highlighting the value of shared resources for advancing new model organisms (The Worm Breeder's Gazette, 2024).

*Data and analysis tool sharing* Furthermore, the community will benefit from tools to share all manner of data imaging datasets, behavioral assays, and functional annotations. Platforms such as the NeuroData Without Borders format for imaging data have enabled open sharing and reuse of data in other systems. In tardigrades, such platforms could facilitate collaborative analysis of high-resolution imaging datasets, genome-wide screens, and standardized behavioral outputs (e.g., locomotion, phototaxis, and stress responses). Developing open-source software for automated imaging, segmentation, and analysis—modeled on existing *C. elegans* tools such as the NeuroPAL pipeline, WormID.org, and behavioral quantification platforms (Javer, 2018)—will accelerate discoveries and lower technical barriers for new labs.

*Community building* Perhaps the largest determinant of success for new model organism introduction is the community of the researchers themselves. Meetings of all sizes—similar to the *C. elegans* or *Drosophila* research meetings—can bring together researchers to share advances, identify priorities, and develop collaborative solutions. The International Symposium on Tardigrada, held every three years, serves as a vital resource for the tardigrade research community. These meetings foster collaboration, highlight cutting-edge techniques, and expand the global understanding of tardigrade biology (International Symposium on Tardigrada, 2024). Establishing additional data and protocol-based open communication platforms, such as a Wiki, forums, or listservs, can enable rapid exchange of ideas and troubleshooting across labs.

In summary, scaling tardigrade research requires:

- ***Bioinformatics databases*** *and tools for genomic data, including **promoters and gene annotations***
- ***Protocol sharing*** *for biological methods*
- *Repositories for **strains and constructs***
- *Open platforms for sharing **datasets** along with software for data processing and visualization.*
- ***Community-building*** *including **meetings** and **collaborative communication channels**, centered on data and protocol development*

## III. COMPREHENSIVE REVIEW OF RELEVANT TARDIGRADE BIOLOGY

### Evolutionary and phylogenetic context

Tardigrades, members of the phylum *Tardigrada*, are microscopic ecdysozoans with an evolutionary position that bridges simple unsegmented organisms, including the nematode *C. elegans*, and arthropods, including *Drosophila* (**Figure 1**A). As part of the Panarthropoda clade, they share a common ancestry with arthropods and onychophorans (velvet worms), offering a critical comparative system for understanding segmentation, appendage evolution, and neurodevelopment (Jørgensen et al., 2018; Doyère, 1842; Marcus, 1929; Smith et al., 2024).

Over 1,200 tardigrade species have been described (Bartels et al., 2016; Nelson et al., 2018) across three classes: Eutardigrada, Heterotardigrada, and Mesotardigrada. These classes are distinguished by differences in morphology, such as the cuticle, buccopharyngeal apparatus, and appendage structures. The most commonly used laboratory strains are eutardigrades (Bartels et al., 2016; Nelson et al., 2018, 2020).

Molecular and fossil evidence situate tardigrades as diverging from their panarthropod relatives during the Cambrian explosion (~500 million years ago), with subsequent evolutionary simplifications that reflect their resilience and compact body plans (Budd, 2001; Mapalo et al., 2024). Unlike arthropods and onychophorans, tardigrades have lost respiratory and circulatory systems, underscoring their evolutionary plasticity and adaptive capacity to thrive in extreme environments (Jørgensen et al., 2018).

Despite their compact size (50–1500 μm), tardigrades retain modular nervous systems, a characteristic shared with other panarthropods. Their nervous system includes a multi-lobed brain, ventral nerve cord, and segmental ganglia, reflecting ancestral traits (Smith et al, 2017) (**Figure 1**B **and 1**C). There is a high degree of molecular conservation of neuronal and neurotransmitter system components (Schumann et al., 2018; Mayer et al., 2015; Dutta et al., 2024; Yamakawa et al., 2024). Advances in molecular phylogenetics and morphological studies have uncovered cryptic species and clarified evolutionary relationships, enhancing their value for comparative research (Jørgensen et al., 2011; Faurby et al., 2011; Guidetti et al., 2009; Gasiorek et al., 2019).

Tardigrades also exhibit diverse reproductive strategies, ranging from bisexual reproduction and hermaphroditism to parthenogenesis, depending on species and environmental pressures (Bertolani, 2001; Guidetti & Bertolani, 2005). Their eggs display remarkable adaptations, including surface sculpturing and encystment, that confer resistance to desiccation and extreme environments (Janelt et al., 2024; Møbjerg et al., 2011; Schill, 2019). These traits enable tardigrades to thrive in diverse habitats and offer insights into how reproductive strategies may intersect with neural and behavioral adaptability.

Tardigrades' evolutionary trajectory reflects a balance of simplification and adaptation, providing an unparalleled opportunity to study the retention, modification, or reduction of neural systems. This phylogenetic context, combined with their reproductive diversity and resilience, positions tardigrades as a vital model for understanding nervous system evolution across ecdysozoans and bridging the gap between nematodes, onychophorans, and arthropods (Moysiuk, 2024; Martin et. al, 2017).

**Physiology and ecology**

The advantages of tardigrades as laboratory organisms extend beyond their phylogenetic position. These microscopic animals are remarkably resilient, thriving in a diverse range of habitats worldwide, from freshwater ponds and marine sediments to semi-terrestrial environments like mosses, lichens, and soil. Their adaptability even extends to extreme conditions, such as deserts, polar ice, and deep-sea trenches, where they survive through cryptobiosis, entering desiccated or frozen states to endure harsh environments (Nelson et al., 2018; Nelson et al., 2020). Their ability to maintain behavioral and physiological processes across such diverse habitats motivates the study in tardigrades of the principles of neural circuit resilience, sensory integration, and adaptive responses to environmental stress (Goldstein, 2022).

The ease of cultivating certain species under controlled laboratory conditions, combined with their ability to survive long-term preservation through desiccation and freezing, makes tardigrades highly versatile for experimental research (**Table 1**) (Roszkowska et al., 2021; Alteiro et al., 2001; Goldstein 2018; Lyons et al., 2024).

**Neuroanatomy**

Electron microscopy, immunohistochemistry and confocal microscopy have found that general organization of the tardigrade nervous system comprises a dorsal brain, ventral longitudinal nerve cords, and paired segmental ganglia (**Table 2**, **Figure 2**) (Wiederhöft and Greven, 1996; Dewel and Dewel, 1996). The brain is typically multi-lobed and bilaterally symmetric, serving as a central hub for sensory integration, and connected to the trunk ganglia by paired longitudinal connectives, with each ganglion providing outputs to corresponding limbs and sensory structures (Gross et al., 2015; Schulze and Schmidt-Rhaesa, 2013). Tools such as anti-acetylated α-tubulin for labeling axonal tracts, anti-synapsin for marking synaptic neuropil, and neurotransmitter markers like serotonin and RFamide have revealed both conserved and species-specific features of the tardigrade nervous system (**Figure 2**B) (Gross and Mayer, 2015; Persson et al., 2012).

X-ray nano-computed tomography (nanoCT) of *H. exemplaris* revealed that the dorsal brain occupies a sizable total body volume (Gross et al., 2019). Remarkably, the tardigrade brain appears to be proportionally larger relative to body size than that of a honeybee and even some larger insects, following the trend described by Haller's Rule (Gross et al., 2019). This suggests an impressive neural investment within a small, efficient body plan.

This anatomically modular neural architecture stands in contrast to the largely non-modular nervous system of *C. elegans*, while retaining an overall neuron count of only a few hundred neurons. As mentioned, the complete neuron count for any tardigrade species has not been determined, and their connectome remains unexplored. Neuronal subtypes—such as motor, sensory, and interneurons—have yet to be definitively identified or characterized. Methods such as in RNA transcript-based *in situ* hybridizations, detailed in [Appendix I](), hold promise for uncovering molecular signatures of neuronal subtypes, aiding efforts to resolve the structure-function relationships of the tardigrade nervous system.

*Comparative Neuroanatomy Across Tardigrade Species* The *H. exemplaris* brain appears unipartite and homologous to the protocerebrum of arthropods (Smith et al., 2018), substantiated by conserved expression patterns analyses of brain-patterning genes such as six3, orthodenticle, and pax6. Immunostaining for β-tubulin has revealed a segmented ventral nerve cord with four paired trunk ganglia, each connected by longitudinal connectives (Gross and Mayer, 2015). These ganglia may be associated with central pattern generators (CPGs) that may coordinate the multi-limbed locomotion characteristic of tardigrades (Nirody, 2021).

In contrast, the neuroanatomy of *Halobiotus crispae*, another eutardigrade, displays a more modular central brain organization, with distinct outer, inner, and median lobes connected by a central neuropil (Persson et al., 2012), and bilaterally symmetric neuronal clusters and transverse commissures within the trunk ganglia.

Heterotardigrades, which include marine species such as *Actinarctus doryphorus* and terrestrial representatives like *Echiniscus testudo*, exhibit additional complexity in their neuroanatomy. In *A. doryphorus*, the brain is tripartite, with distinct lobes dedicated to innervating cephalic sensory structures (Persson et al., 2014), highlighting the modularity and sensory specialization seen in marine tardigrades. Similarly, in *E. testudo*, the brain features a prominent horseshoe-shaped neuropil with two major commissures and ventrolateral clusters innervating cephalic sensory fields (Gross et al., 2021; Schulze and Schmidt-Rhaesa, 2013). Anti-synapsin staining has been particularly effective in mapping synaptic organization, revealing how compact neural circuits integrate multimodal sensory information.

Ventral nerve cord organization is remarkably consistent across tardigrades, with paired ganglia housing bilaterally symmetric neurons. Commissural fibers link the ganglia, forming connections that allow for both local and global coordination of motor outputs. In *Macrobiotus cf. harmsworthi* and *Paramacrobiotus richtersi*, anteriorly shifted ganglia relative to their limbs suggest an evolutionary adaptation for compact neural processing (Mayer et al., 2013; Zantke et al., 2008). Further studies using allatostatin-like and perisulfakinin-like immunostaining have revealed a diversity of neurotransmitter systems operating within the trunk ganglia (Persson et al., 2012).

More recent studies have also examined how the nervous system is preserved or reorganized during dormancy. In *Thulinius ruffoi*, an encysting eutardigrade, a combination of light microscopy, transmission electron microscopy (TEM), and serial block-face scanning electron microscopy (SBEM) revealed the persistence of neural clusters in the cephalic region and ventral nerve cord during prolonged encystment (Janelt and Poprawa, 2024). Neuronal structures, including somata and synaptic regions, remain intact, with minimal degradation over extended periods.

**Sensory organs**

While compact, tardigrade nervous systems feature a remarkably versatile and specialized array of sensory structures that enable them to navigate and adapt to diverse environments. Their sensory systems detect visual, mechanical, and chemical cues, allowing these microscopic animals to perform context-specific behaviors such as phototaxis, substrate navigation, and feeding—although the physiology and mechanisms of these structures is little-understood.

*Photoreception and light perception* Tardigrade photoreception is likely effected by two visible eyespots located on the anterior dorsal surface of the head. These eyespots, consisting of pigment cups paired with photoreceptive cells, seem to be homologous to the median ocelli of other panarthropods (Martin et al., 2017; Hering and Mayer, 2014). While rudimentary compared to the complex eyes of arthropods, the visual system of tardigrades is functionally significant (Greven, 2007). Recent studies have identified multiple rhabdomeric opsins (r-opsins) and ciliary opsins in tardigrades, including lineage-specific duplications, suggesting an evolutionary diversification of light perception mechanisms (Hering and Mayer, 2014; Fleming et al., 2021). In *H. exemplaris*, differential expression of opsins across life stages indicates that photoreceptive systems vary throughout development, likely reflecting changing ecological demands (Fleming et al., 2021). Although tardigrades are unlikely to distinguish colors due to a lack of co-temporal opsin expression, their photoreceptive system remains functionally versatile. Opsins may contribute not only to phototaxis but also to circadian regulation and other light-dependent processes (Fleming et al., 2021). Recent work has demonstrated the presence of pigment-dispersing factor (PDF) neuropeptides, which act as neuromodulators connecting light detection to physiological and behavioral rhythms, such as locomotion and feeding (Dutta et al., 2024).

*Mechanoreception and tactile sensory structures* Mechanosensation in tardigrades is believed to be mediated by sensory fields and appendages distributed across the head and body. Prominent cephalic structures, including cirri and clavae, are likely innervated by sensory neurons that connect to the brain. Cirri are filament-like appendages that likely function as tactile sensors, while clavae, club-shaped projections, may also detect mechanical and chemical stimuli (Biserova and Kuznetsova, 2012; Møbjerg et al., 2018). Transmission electron microscopy (TEM) studies in *Halobiotus stenostomus* reveal the presence of ciliated receptor endings within the sensory fields, with structural diversity suggesting functional specialization in detecting mechanical deformation or substrate vibrations (Biserova and Kuznetsova, 2012). Mechanosensory inputs are likely further detected by structures within the limbs and claws, which allow tardigrades to interact with and navigate their surroundings (Suzuki, 2022; Nelson et al., 2018). The specialized claws at the tips of their limbs seemingly provide tactile feedback that integrates with motor outputs for substrate adhesion and multi-limbed gait coordination. These putative mechanoreceptors are likely critical for tardigrades as they navigate complex substrates such as moss, sediment, and algae (Nirody et al., 2021).

*Chemosensation and Suboral Sensory Organs* Tardigrades also exhibit chemosensory abilities, likely mediated by suboral sensory structures and associated fields located near the buccal cavity. The pharyngeal organ, found in species such as *M. tardigradum*, contains ciliated receptor pockets thought to detect food-related chemical cues within the buccal tube (Dewel and Clark, 1973; Guidetti et al., 2012). Similarly, studies in *H. stenostomus* have identified specialized receptor endings in the suboral region, which may play a role in chemosensory detection (Biserova and Kuznetsova, 2012). These sensory systems likely enable behaviors such as chemotaxis, where tardigrades respond to chemical gradients in their environment to locate food or avoid harmful substances.

The cephalic sensory organs—including the eyespots, cirri, and clavae—anatomically apear to connect to the brain's multi-lobed structure, which serves as a hub for processing visual, mechanical, and chemical information. Simultaneously, tactile sensory fields and mechanosensors within the limbs may provide essential feedback for motor control, enabling precise coordination of multi-limbed locomotion.

While tardigrade sensory systems have been described to an extent at the anatomical and morphological levels, our understanding of their functional mechanisms remains limited. Current insights into how tardigrades perceive and process sensory inputs are primarily inferred from structural observations—such as the presence of photoreceptors, ciliated mechanosensory cells, and cephalic sensory fields—rather than direct functional evidence. For instance, while multiple opsin genes have been identified, the pathways by which light signals are transduced, integrated, and transformed into behavioral outputs like phototaxis remain unknown (Hering and Mayer, 2014; Fleming et al., 2021). Similarly, mechanoreceptive and chemosensory structures, though well-documented at ultrastructural levels, lack functional characterization.

**Behavior**
*Limbed locomotion* Tardigrades have eight legs in four pairs and display a distinctive locomotion style that combines simplicity with surprising sophistication. Tardigrade gait patterns and biomechanics share remarkable similarities with much larger arthropods, despite significant differences in scale, body composition, and habitat (Nirody et al., 2021; Nirody, 2021). High-speed video tracking shows that limno-terrestrial tardigrades primarily utilize a tetrapod-like gait, where limbs swing in a coordinated back-to-front wave along each side of the body (Nirody et al., 2021; Shcherbakov et al., 2010). This stepping pattern aligns with local coordination rules observed in insects, where the stance-to-swing transition in one leg is suppressed if a neighboring ipsilateral leg is already in swing phase and promoted when the neighboring leg completes its stance phase (Cruse's Rules; Cruse, 1990; Nirody et al., 2021). These rules give rise to smooth transitions between stepping patterns as walking speed increases, without the abrupt gait shifts seen in vertebrates. Instead, tardigrades exhibit a continuum of inter-limb coordination patterns (ICPs), modulating stance duration while keeping swing duration relatively constant, a hallmark of arthropod locomotion strategies (Nirody, 2021). These rules, along with largely uncharacterized sensory feedback mechanisms such as proprioception and load sensing, may allow tardigrades to adapt their stepping patterns dynamically, even when navigating uneven or unstable substrates (Zill et al., 2011; Nirody, 2023).

Biomechanical studies reveal that tardigrades' walking kinematics are robust across speeds and environmental challenges. For example, when walking on softer substrates, such as low-stiffness gels, H. exemplaris adapts by transitioning to a "galloping" coordination pattern, where ipsilateral legs swing more synchronously, improving locomotion efficiency on deformable surfaces (Nirody et al., 2021; Anderson et al., 2024). Similarly, tardigrades walking on uneven or three-dimensional terrains demonstrate remarkable coordination, aligning with behaviors observed in other panarthropods like stick insects and spiders when traversing complex environments (Dürr et al., 2018; Nirody, 2023). This continuum reflects a probabilistic modulation of coordination patterns, a feature shared across panarthropods, highlighting the flexibility and adaptability of tardigrade locomotion even at varying speeds (DeAngelis et al., 2019; Nirody, 2023). Tardigrades' locomotion is not only impressive in its efficiency but also in its resilience to environmental perturbations. *M. tardigradum* shows adaptability of both speed and direction in response to environmental cues, exhibiting behaviors like orthokinesis and klinokinesis to remain in favorable microhabitats or escape desiccation (Shcherbakov et al., 2010).

The fourth pair of posterior limbs in tardigrades also displays apparent functional specialization or versatility. Unlike the anterior three pairs, the fourth pair is often used for grasping and anchoring, especially when navigating complex three-dimensional environments (Nirody et al., 2021; Anderson et al., 2024). However, kinematic analyses suggest that these posterior limbs are still incorporated into the overall coordination scheme during locomotion, albeit with reduced stepping precision (Anderson et al., 2024).

Tardigrade locomotion studies primarily focus on limno-terrestrial species like *H. exemplaris* and *M. tardigradum*, which traverse substrates such as moss or agarose gels using their limbs for walking and climbing (Shcherbakov et al., 2010; Anderson et al., 2024). In contrast, marine tardigrades use distinct strategies like swimming aided by sensory appendages, highlighting the biomechanical diversity across the clade (Nelson et al., 2018) and suggest comparative studies of functional adaptations to different environments.

The neural basis of tardigrade locomotion has not been studied. In particular, the neural mechanisms that coordinate limb movements to produce gaits, or facilitate transitions between gaits, are unknown.

*Sensory Navigation* Tardigrades exhibit a diverse range of goal directed behaviors, including navigation of light, temperature, and chemical gradients, modulation of activity states, and conditioned learning. Phototaxis, the ability to move in response to light, has been observed in several species, where they often exhibit positive or negative responses depending on life stage or environmental conditions (Beasley, 2001; Greven, 2007; Meyer et al., 2020). For example, predator-prey interactions reveal that tardigrades can detect chemical cues: *Milnesium lagniappe*, a predatory species, is attracted to areas previously occupied by its prey, while prey species like *Macrobiotus acadianus* actively avoid such regions, likely relying on olfactory or chemical signals rather than vision (Meyer et al., 2020). Similarly, sex-specific responses to waterborne chemical cues have been demonstrated in *Macrobiotus polonicus*, where males preferentially approach females, highlighting the role of chemical communication in mate searching (Chartrain et al., 2023). The molecular and neural basis of sensory perception and neural signal integration remains largely unexplored (Dewel and Clark, 1973; Hvidepil and Møbjerg, 2023).

*Activity Modulation and Dormancy* Tardigrades exhibit remarkable behavioral flexibility in response to environmental stress. The most well-known form of this is anhydrobiosis, a reversible dormant state induced by desiccation, where tardigrades retract their limbs into a "tun" and drastically reduce metabolic activity to endure prolonged stress (Hvidepil and Møbjerg, 2023). Similar behavioral responses, including osmobiosis and chemobiosis, are observed under osmotic stress and exposure to toxicants (Hvidepil and Møbjerg, 2023). Interestingly, tardigrades can enter reversible paralysis-like states under less extreme conditions. For instance, cold exposure induces a cold-induced coma, a behavioral state where motor activity ceases but is rapidly restored upon warming, indicating an ability to maintain neural and muscular integrity (Lyons et al., 2022).

*Learning and Memory* Tardigrades appear to exhibit basic forms of learning, such as aversive conditioning. *Dactylobiotus dispar* associated a neutral stimulus (blue light) with an aversive electrical shock, displaying a curling defensive behavior when subsequently exposed to the light alone (Zhou et al., 2019). This short-term memory, established after a single pairing, highlights the potential of tardigrades as a model to study one-shot learning. Given their ability to enter metabolic dormancy, tardigrades offer a novel system for exploring how memories are maintained or disrupted during periods of low or absent metabolic activity.

**Genomics resources and genetic engineering**

*Genomics* *Hypsibius exemplaris* (formerly known as *H. dujardini*) and *Ramazzottius varieornatus* were the first tardigrade species to receive high quality genome assemblies (Yoshida et al., 2017). Their genomes, ranging from 56 Mb (*R. varieornatus*) to 104 Mb (*H. exemplaris*), encode 13,000–19,000 genes, a count similar to nematodes and insects, yet with streamlined intergenic regions and shorter introns (Yoshida et al., 2017). This genomic simplicity reduces redundancy, making tardigrades particularly relevant for identifying clear relationships between genes, circuits, and behaviors. The diploid nature of their genomes, with 2n = 10 in H. exemplaris and 2n = 12 in *R. varieornatus*, further enhances their accessibility for genetic studies (Ammermann 1967; Yoshida et al., 2017). An early claim of extensive horizontal gene transfer (HGT) in tardigrades, which suggested that foreign bacterial genes could explain their extremotolerance, has since been refuted (Boothby et al., 2015). Current analyses place HGT at less than 0.4% of their genomes—comparable to other eukaryotes—following improvements in sequencing methods and contamination controls (Richards and Monier, 2016; Koutsovoulos et al., 2016; Arakawa et al., 2016).

Other tardigrade genomes have since been sequenced, expanding the genomic resources for comparative analyses. The genome of *Echiniscus testudo*, a heterotardigrade, has provided further insight into tardigrade body plan evolution and gene loss events, such as the absence of certain *Hox* genes, that underpin their simplified morphology (Murai et al., 2021; Smith et al., 2024). Similarly, *Paramacrobiotus metropolitanus sp. nov.*, *Paramacrobiotus richtersi*, and *Milnesium tardigradum* genomes have revealed important lineage-specific adaptations and expanded our understanding of tardigrade diversity and resilience and body plan evolution (Sugiura et al., 2022; Smith et al., 2024; Bemm et al., 2017). Notably, the genome of *Hypsibius henanensis*, a newly identified species, has been sequenced, providing insights into the molecular basis of radiotolerance, including the identification of a horizontally transferred gene, Doda1, which produces protective pigments called betalains, and another gene, TRID1, involved in DNA repair (Li et al., 2024). These additional genomes underscore the evolutionary plasticity of tardigrades and provide a broader comparative framework to explore how their streamlined genetic architecture supports their unique biology.

*Early progress in genetic engineering - RNAi* The development of genetic tools for manipulating tardigrade genomes is in its infancy. RNA interference (RNAi) was the first tool applied to disrupt gene function in tardigrades, where RNAi delivered by microinjection in adult in *H. exemplaris* successfully targeted developmental genes like *mago-nashi*, a gene required for embryonic elongation (Tenlen et al., 2013). Knockdown of *mago-nashi* resulted in observable embryonic phenotypes, such as failure to elongate during development. Other axial patterning genes were also targeted, producing effects largely consistent with those seen in other model organisms. However, RNAi efficiency varies, and subtle or absent phenotypes require transcript-level validation for confirmation, limiting reliability for functional studies.

*CRISPR-based methods* The advent of CRISPR/Cas9 genome editing marked an important step forward for many non-model organisms, but has yet to overcome critical hurdles in tardigrades. Early attempts involved microinjection of Cas9 ribonucleoprotein complexes directly into the body cavities of hundreds of adult *H. exemplaris* (Kumagi et al., 2022). These microinjections—technically demanding and time-intensive—produced somatic mutations that required genotyping to verify. No observable phenotypes were reported, and editing efficiency remained exceptionally low, highlighting the limitations of this approach for generating functional data at the level of other sister taxa and standard model organisms. A recent advance was the development of DIPA-CRISPR in *R. varieornatus*, which enables heritable gene editing by injecting Cas9 RNPs into parental females during vitellogenesis (Kondo et al., 2024). DIPA-CRISPR produced homozygous mutants for genes like RvLEAM and tubulin in a single step (still requiring laborious microinjections into adult body cavities), but phenotypic changes were limited to molecular verification, with no gross morphological or behavioral phenotypes observed (Kondo et al., 2024). Additionally, while knock-in experiments have been demonstrated, they are currently restricted to short insertions of a few base pairs; no study has yet successfully integrated larger constructs, such as fluorescent reporters or entire genes. These shortcomings limit the utility of CRISPR for scalable genetic manipulation and the generation of stable tools critical for neuroscience applications.

*Extrachromosomal vectors* Transient gene expression and the visualization of protein dynamics has been recently demonstrated in living tardigrades (Tanaka et al., 2023), dubbed the TardiVec system. TardiVec vectors utilize 1-kilobase (kb) promoter sequences extracted from tardigrade genomes, particularly from highly expressed housekeeping genes such as actin, tubulin, and ubiquitin (Tanaka et al., 2023). This promoter length was selected based on the compact nature of tardigrade genomes, where regulatory sequences are likely confined to short upstream regions. TardiVec has successfully driven the expression of fluorescent reporters such as GFP and GCaMP6s across a variety of tissues, including muscles, the epidermis, storage cells, and the nervous system. Beyond protein visualization, the system can be effective for promoter screening, allowing researchers to test and validate the regulatory activity of native sequences with precision. TardiVec is currently limited to transient expression and delivery of plasmids relies on electroporation, which shows variability in transgene uptake across tissues and individuals. Over time, fluorescent signals diminish as the plasmid degrades, restricting its application for long-term studies or stable lineage tracing (Tanaka and Kunieda, 2023).

## IV. WHICH SPECIES OF TARDIGRADE FOR SYSTEMS NEUROSCIENCE?

While *Hypsibius exemplaris* is the frontrunner for systems neuroscience research, thanks to its transparency, ease of culture, published genome, and early attempts at transgenic methods that include transient expression of GCaMP6s, other species bring unique strengths to the field (Yoshida et al., 2017; Tanaka et al., 2023). For instance, *Ramazzottius varieornatus* is renowned for its extraordinary resistance to desiccation and radiation, alongside its wealth of publicly accessible genomics resources, making it a valuable model for investigating the molecular mechanisms of stress resilience, yet has pigmentation in its body cuticle that may interfere with imaging of neural indicators (Yoshida et al., 2017; Ishikawa et al., 2024; Horikawa et al., 2013; Horikawa et al., 2012; Horikawa, 2008; Hashimoto et al., 2016). *Milnesium tardigradum*, a predatory tardigrade, offers insights into behavioral adaptations and evolutionary diversity, while *Paramacrobiotus metropolitan sp. nov*. or *Richtersius coronifer* could be further developed as genetic models that sexually reproduced (can could be crossed) and has a larger body size (which may aid in downstream neural imaging, if transgenic strains can be generated) (Bemm et al., 2017; Sugiura et al., 2022; Stec et al., 2020). Each species provides unique opportunities to study tardigrade biology under different experimental conditions.

Cultures of *H. exemplaris* are particularly well-suited for laboratory research due to their simple maintenance requirements and predictable life cycle (Goldstein, 2022). This species thrives in freshwater media, such as commercially sourced mineral water, and feeds on algal species like *Chlorella vulgaris* or *Chlorococcum sp*. (Goldstein, 2018; Lyons et al., 2024). Under low light conditions at 17-20°C, *H. exemplaris* completes its life cycle in approximately 7–14 days, depositing ~1-8 embryos in synchronous clutches during molting. These characteristics enable rapid generational turnover (though much slower than *C. elegans*) and scalability for experimental studies, making *H. exemplaris* an excellent candidate for high-throughput approaches such as transcriptomic analysis, behavioral assays, and imaging-based workflows.

Another useful feature of *H. exemplaris* is its transparency throughout all life stages, which facilitates high-resolution visualization of fluorescent signals and internal structures (Gross and Mayer et al., 2015; Tanaka and Kanieda, 2023; Bartels et al., 2023). This optical clarity is a major advantage for neuroscience research, allowing non-invasive imaging of cellular and molecular processes such as calcium dynamics, gene expression, and stress responses. While other species like *R. varieornatus*, *M. tardigradum, and E. sigismundi* may contribute unique insights based on their differentiated stress tolerance and behavior, they contain pigmentation in their body cuticle that may interfere with live imaging (Horikawa et al., 2008; Suzuki 2003; Hvidepil and Møbjerg, 2023).

In the context of systems neuroscience, *H. exemplaris* remains the most practical and experimentally versatile species, and several neuroanatomical studies have already been completed on *H. exemplaris*. Nevertheless, the inclusion of other tardigrade species in experimental research will broaden the scope of potential discoveries, particularly in comparative neurophysiology, ecological adaptation, and stress tolerance.

## CONCLUSION

Tardigrades offer a unique combination of experimental accessibility, evolutionary relevance, and behavioral richness that positions them as a transformative model for systems neuroscience. However, realizing their potential requires a concerted effort to develop the necessary tools and resources. By addressing needs in molecular biology, neuroanatomy, imaging, and behavioral assays, we can unlock the full power of tardigrades as a model organism. Achieving these goals will not only bridge critical gaps in neuroscience but also expand our understanding of how compact nervous systems generate, adapt, and optimize behavior across diverse conditions. The time to invest in tardigrade neuroscience is now, and the rewards promise to be transformative.

## AUTHOR CONTRIBUTIONS

**A.M.L**: Conceptualization (equal), Data Curation (lead), Formal Analysis (lead), Funding Acquisition (equal), Visualization (lead), Writing – Original Draft Preparation (lead), Writing – Review & Editing (equal). **S.K.**: Conceptualization (equal), Data Curation (supporting), Funding Acquisition (equal), Visualization (supporting), Writing – Review & Editing (equal).

## DECLARATION OF COMPETING INTERESTS

The authors declare no conflict of interest. The funders had no role in the writing of the manuscript, or in the decision to publish the findings.

## FUNDING

Our work was partially supported by a National Science Foundation Postdoctoral Research Fellowships in Biology (PRFB) Award (2109906, A.M.L.), the Grass Foundation Fellows Program (2024, A.M.L.), and the Weill Institute of Neuroscience.

## ACKNOWLEDGEMENTS

Frank Smith, Ph.D. (University of California, Berkeley), Mandy Game (University of North Florida), Heather Bruce Ph.D. (University of North Florida), Nipam Patel Ph.D (Marine Biological Laboratory, University of Chicago), and Guilherme Gainett, Ph.D. (Boston Children's Hospital and Harvard University) provided helpful suggestions for optimizing HCR RNA *in situ* conditions, of previously unpublished data shown in Figure 2B. Frank Smith, Ph.D. generously provided the HCR probe design for the *elav* transcript visualized (Molecular Instruments, Inc.) in Figure 2B. Carsten Wolff, Ph.D. (Marine Biological Laboratory, University of Chicago) aided in optimizing microscopy conditions for Figure 2B. Hannah Gemrich (University of California, San Francisco) produced the immunohistochemistry image generously provided for use in Figure 2C. Yuya Yoshida generously provided artistic 3D renderings of *H. exemplaris* in Figure 4B and Figure 5B. We thank all future colleagues and community members for helpful comments in future iterations of this manuscript.

# Tables

**Table 1. Overview of Neuroanatomical Studies in Tardigrades: Species, Techniques, and Key Findings** A summary of tardigrade species with neuroanatomical studies, including their habitats, tools and techniques employed (e.g., electron microscopy, immunohistochemistry), antibodies/stains used, resolution achieved for neuronal mapping, and significant findings.

| Species | Habitat | Tools/Techniques | Antibodies/Stains Used | Resolution for Segmenting Neurons or Identifying Cells | Lab Strain/Genome | Key Findings | Citation |
|---|---|---|---|---|---|---|---|
| *Echiniscus viridissimus* | Terrestrial | Electron Microscopy (EM), Contrast Agents | Contrast Agents (Uranyl Acetate, Lead Citrate) | Ultrastructural resolution; neuronal clusters and sensory ganglia | No | Multi-lobed brain with ganglia innervating cirri and clavae | Dewel and Dewel (1996) |
| *Milnesium tardigradum* | Terrestrial | Electron Microscopy (EM), Contrast Agents | Contrast Agents (Uranyl Acetate, Lead Citrate) | Ultrastructural resolution; neuronal clusters and tracts | Yes; genome published | Saddle-shaped brain with outer and inner lobes; cephalic sensory inputs | Wiederhöft and Greven (1996) |
| *Macrobiotus cf. harmsworthi* | Terrestrial | Immunohistochemistry, Confocal Microscopy, Nuclear Stains | Acetylated α-tubulin, Serotonin, RFamide, Hoechst (nuclei) | Moderate resolution; ganglia, anteriorly shifted neurons, nuclear staining | No | Segmental ganglia with transverse commissures; anteriorly shifted ganglia | Zantke et al. (2008), Mayer et al. (2013) |
| *Halobiotus crispae* | Marine/ Brackish | Immunohistochemistry, Confocal Microscopy, Nuclear Stains | Acetylated α-tubulin, Serotonin, RFamide, Hoechst (nuclei) | Moderate resolution; bilaterally symmetric neurons and commissures | No | Modular brain with distinct lobes and central neuropil; bilateral symmetry | Persson et al. (2012) |
| *Paramacrobiotus richtersi* | Terrestrial | Immunohistochemistry, Confocal Microscopy | Allatostatin-like, Perisulfakinin-like | Moderate resolution; neurotransmitter systems in ganglia | No | Neurotransmitter diversity in trunk ganglia; sensory-motor processing | Persson et al. (2012) |
| *Actinarctus doryphorus* | Marine | Immunohistochemistry, Confocal Microscopy | Acetylated α-tubulin | Moderate resolution; brain lobes and cephalic sensory structures | No | Tripartite brain with sensory specialization for cephalic appendages | Persson et al. (2014) |
| *Hypsibius exemplaris* | Freshwater | Immunohistochemistry, Confocal Microscopy, HCR/in situ Hybridization, Nuclear Stains | Anti-β-tubulin, Acetylated α-tubulin, HCR probes, Hoechst (nuclei) | Moderate resolution; segmented ventral nerve cord and trunk ganglia, gene expression patterns | Yes, established lab strain; genome published | Unipartite brain homologous to arthropod protocerebrum; modular ganglia | Smith et al. (2018), Gross and Mayer (2015) |
| *Hypsibius exemplaris* | Freshwater | MicroCT, NanoCT Imaging | OsO4, Ferrocyanide, Critical Point Drying | High resolution; 3D segmentation of brain and trunk ganglia | Yes, established lab strain; genome published | 3D reconstruction of brain and ganglia; spatial relationships and volumes quantified | Gross et al. (2019) |
| *Echiniscus testudo* | Semi-terrestrial | Immunohistochemistry, Confocal Microscopy, Nuclear Stains | Anti-synapsin, Acetylated α-tubulin, DAPI (nuclei) | High resolution; horseshoe neuropil, ventrolateral clusters, nuclear organization | No | Horseshoe-shaped neuropil; commissures connecting sensory inputs | Gross et al. (2021), Schulze and Schmidt-Rhaesa (2013) |
| *Thulinius ruffoi* | Freshwater | Light Microscopy, TEM, SBEM, Nuclear Stains | Methylene Blue, TEM contrast agents, Acetylated α-tubulin, Hoechst (nuclei) | Moderate resolution; cephalic clusters, ventral nerve cord persistence, nuclear staining | No | Persistent cephalic clusters and ventral nerve cord during dormancy | Janelt and Poprawa (2023) |

**Table 2. Characteristics of emerging tardigrade species** classified by their model status (widespread, moderate, or developing), with details provided on genome availability, genome size, reproductive mode, habitat, optical properties, average body size, neuroscience applications, and associated challenges for research. All species have limited genetic tools and protocols, although *H. exemplaris* and *R. varieornatus have* been at the forefront of genetic engineering methods development.

| Species | Genome Published | Genome Size | Reproductive Mode | Habitat | Optical Properties | Average Body Size (microns) | Neuroscience Applications | Challenges for Research |
|---|---|---|---|---|---|---|---|---|
| *Hypsibius exemplaris* | Yes | 104.4MB, 18,475 genes (Yoshida et al., 2017) | Parthenogenetic | Freshwater, moss, and soil environments | Transparent | 200-300 | Tractable for genetic manipulation and neural imaging | Small cell size & neural structures may be difficult to image |
| *Ramazzottius varieornatus* | Yes | 54.2MB, 13,000 genes (Yoshida et al., 2017) | Parthenogenetic | Freshwater and terrestrial environments, often in mosses or lichens | Pigmented (reddish-brown) | 300-500 | Many resilience mechanisms linked to stress tolerance | High pigmentation complicates optical imaging |
| *Paramacrobiotus metropolitanus* | Yes | 170MB (Hara et al., 2021) | Sexual | Terrestrial habitats, including moss and leaf litter | Transparent | 300-700 | Sex-biased gene expression linked to dimorphism and stress tolerance | Sexual reproduction and male-biased genes complicate standardization |
| *Milnesium tardigradum* | Yes | 75.1MB, 19,401 genes (Bemm et al., 2017) | Sexual | Freshwater and terrestrial environments, including moss and soil | Pigmented (orange-reddish) | 300-700 | Potential for studying predatory behaviors | Carnivorous diet and large size complicate maintenance |
| *Hypsibius henanensis sp. nov.* | Yes | 112.6MB, 14,701 genes (Li et al., 2024) | Likely parthenogenetic | Terrestrial mountainous environments, often in mosses or lichens | Transparent | ~300 | Radiation resistance, including DNA repair and antioxidant pathways | Limited genetic tools and protocols |
| *Macrobiotus shonaicus* | - | Unknown | Sexual | Freshwater and terrestrial habitats, especially mosses and leaf litter | Transparent | 300-500 | Intermediate body size and transparency | Sexual reproduction may complicate strain maintenance |
| *Paramacrobiotus richtersi* | - | Unknown | Sexual and parthenogenetic | Terrestrial environments, mosses, and leaf litter | Transparent | 200-400 | Adaptations to extreme environments | Limited genetic tools and protocols |
| *Richtersius coronifer* | - | Unknown | Sexual | Terrestrial environments, often in moss and soil | Transparent | 500-1000 | Large body size may aid neural recordings | Limited genetic tools and protocols |
| *Halobiotus crispae* | - | Unknown | Hermaphroditic | Marine and brackish water environments | Transparent | 200-300 | Marine sensory adaptations | Requires precise marine conditions for culturing |
| *Echiniscoides sigismundi* | - | Unknown | Likely sexual | Marine and intertidal environments, often near tidal pools | Pigmented (brown) | 300-500 | Adaptations to fluctuating salinity | Marine conditions and pigmentation hinder imaging |
| *Dactylobiotus dispar* | - | Unknown | Parthenogenetic | Freshwater habitats, including ponds and temporary water bodies | Transparent | 250-400 | Potential for studying freshwater adaptations; forms cysts as part of life cycle | Limited tools and baseline data |
| *Acutuncus antarcticus* | - | Unknown | Parthenogenetic | Antarctic freshwater environments, often in mosses and algae | Transparent | 300-500 | Cold adaptation mechanisms | Sensitive to high temperatures, genome editing may be hindered |

**Table 3. Overview of Currently Recognized Sensory Modalities in Tardigrades**
Known or putative structures involved, identified molecular markers where available, and key references. While aspects of photoreception, mechanoreception, and chemoreception have been documented, thermoception remains largely inferred from indirect behavioral changes. Many sensory modalities and underlying mechanisms likely remain undiscovered in the tardigrade.

| Sensory Modality | Known/Putative Sensory Structures | Associated Molecular Markers | Key References |
|---|---|---|---|
| *Photoreception* | Eyespots (anterior dorsal pigment cups, photoreceptor cells) | Opsin genes (r-opsins, c-opsins); PDF neuropeptides | Hering and Mayer, 2014; Fleming et al., 2021; Dutta et al., 2024 |
| *Mechanoreception* | Cirri, clavae, ciliated mechanosensory cells, limb claws | None explicitly identified at molecular level (structural indicators: ciliated receptor endings) | Biserova and Kuznetsova, 2012; Guidetti et al., 2012; Møbjerg et al., 2018 |
| *Chemoreception* | Suboral sensory fields, pharyngeal organ with receptor pockets | None explicitly identified; likely specialized receptor endings or chemoreceptor proteins | Dewel et al., 1993; Biserova and Kuznetsova, 2012; Møbjerg et al., 2018 |
| *Thermoception* | Not clearly identified; responses inferred from behavioral changes (e.g. thermal preferences, cold-induced coma) | No known molecular markers | Li and Wang, 2005; Hvidepil and Møbjerg, 2023; Lyons 2022; Lyons et al., 2024 |

# Figures

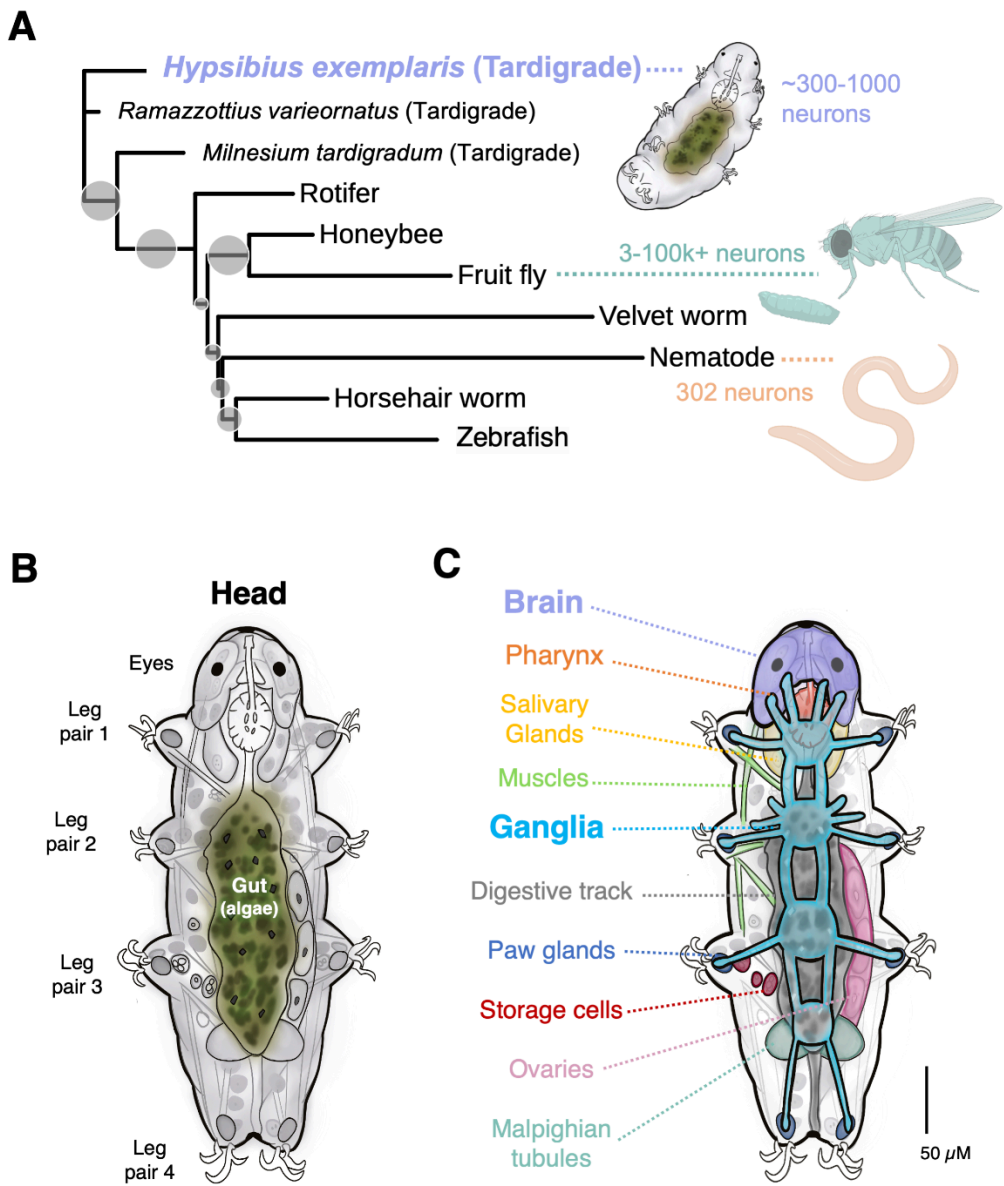

**Figure 1. Tardigrades as an emerging model for systems-level neuroscience**

(A) Phylogenetic placement of emerging model tardigrade species (*H.exemplaris*, *R. varieornatus*, and *M.tardigradum*) among other invertebrates and systems neuroscience models. The maximum likelihood (ML) tree was constructed using 18S ribosomal RNA sequence data, aligned in MUSCLE and analyzed in RAxML with 100 bootstrap replicates. Grey circles indicate bootstrap support, with sizes ranging from 28 (smallest circle) to 100 (largest circle). Comparative neuron counts are shown for organisms commonly used in neuroscience, including the nematode (302 neurons) and fruit fly (100k+ neurons in adults, 3k in larvae). (B) Anatomical diagram of a tardigrade (*H. exemplaris*), depicting external and internal structures such as eyes, legs, and gut (often containing algae from feeding), as visible from bright-field microscopy.
(C) Detailed internal anatomy of *H. exemplaris*, showing key organ systems relevant to neuroscience, including the brain, ganglia, muscles, and digestive tract. The branching of ganglia to limbs is speculative, based on unpublished RNA in situ data and historical anatomical diagrams. The centralized brain and modular ganglia organization suggest organization for distributed motor control. Illustration adapted from nanoCT data (Gross et al., 2017).

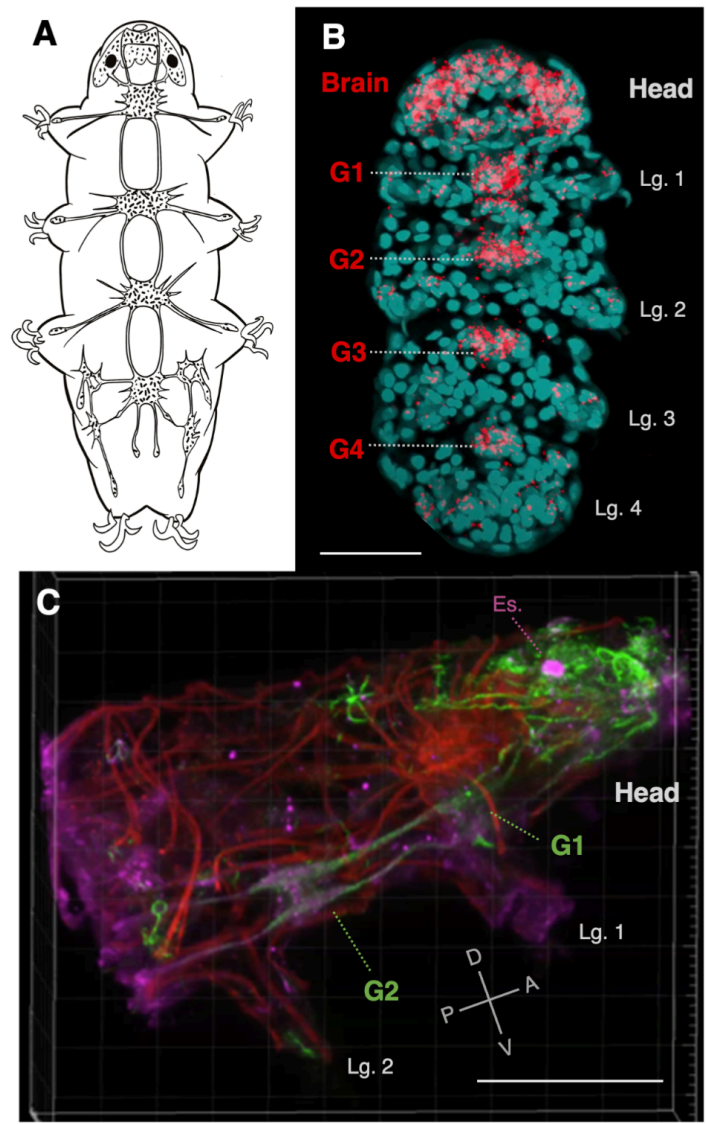

**Figure 2. Historical and modern visualizations of the tardigrade nervous system using light microscopy, *in situ* hybridization, and immunohistochemistry**

(A) Illustration of the nervous system of an example tardigrade species, based on light microscopy, as depicted in historical studies (early 1890s). Ventral view of the tardigrade nervous system, showing the modular arrangement of ganglia and branching to the limbs.

(B) Fluorescence confocal image of a *Hypsibius exemplaris* hatchling showing RNA in situ hybridization (HCR) signal of pan-neuronal *elav* transcript (red), with nuclei stained using DAPI (cyan). Ganglia (G1–G4) correspond to leg pairs (Lg. 1–4). Scale bar: ~25 µm. RNA *in situ* hybridization is easily-customizable and provides qualitative, spatial gene expression patterns.

(C) Fluorescence confocal image of the head region of *H. exemplaris*, visualizing a subset of the nervous system and musculature using immunohistochemistry. Red: F-actin (muscle-like structures); green: Beta-tubulin iso1; magenta: horseradish peroxidase. D: dorsal; V: ventral; A: anterior; P: posterior. Scale bar: ~50 µm. Immunohistochemistry provides high-resolution protein localization but is limited by the availability of specific antibodies and the potential for cross-reactivity.

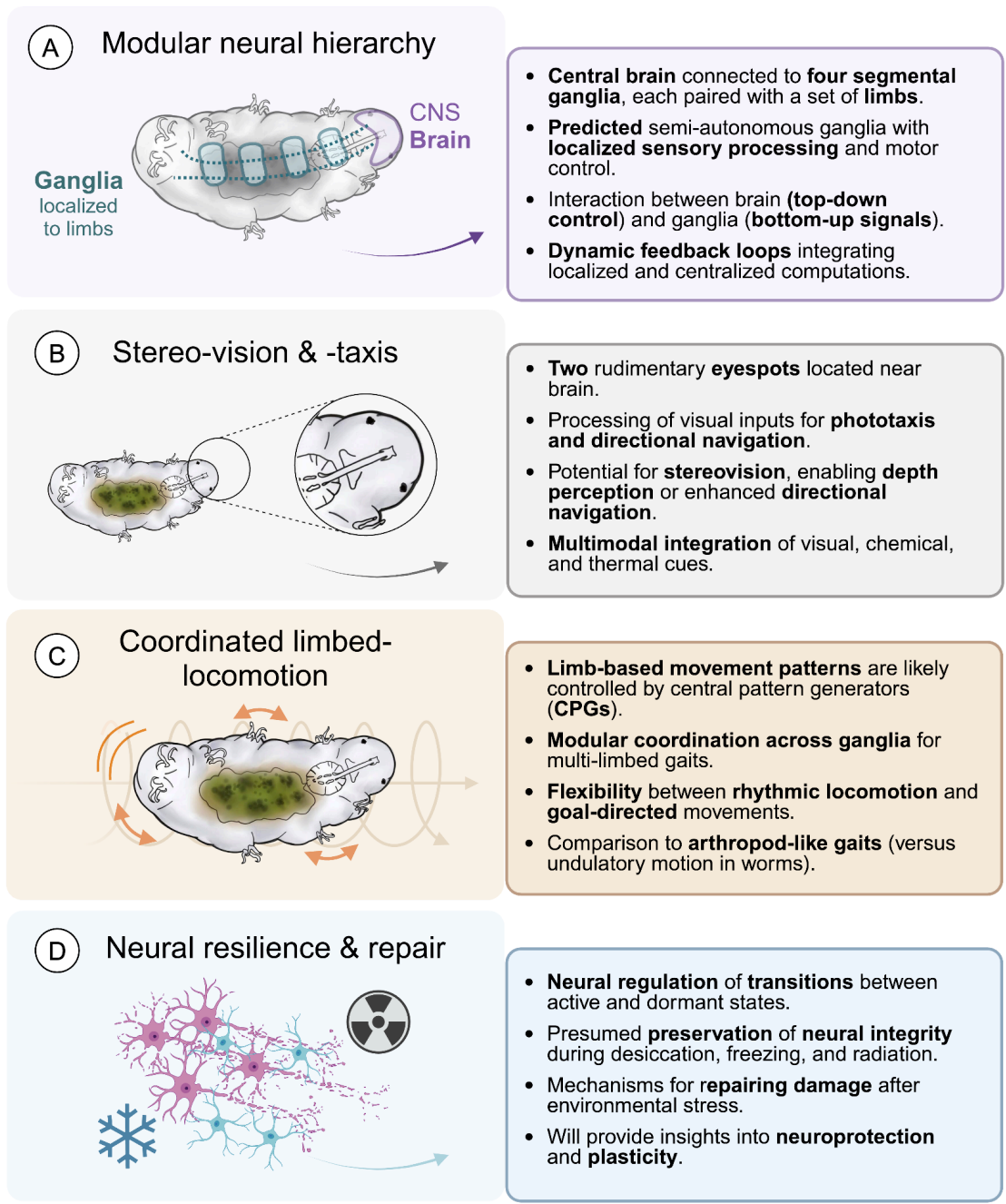

**Figure 3. Scientific rationale to develop tardigrades for systems neuroscience**

(A) Modular neural hierarchy with a central brain and segmental ganglia supporting sensory-motor integration through dynamic feedback loops.
(B) Stereo-vision and -taxis driven by rudimentary eyespots for phototaxis, directional navigation, and multimodal sensory integration.
(C) Coordinated limbed-locomotion regulated by central pattern generators (CPGs), enabling flexible, goal-directed gaits.
(D) Neural resilience and repair mechanisms, preserving neural integrity during desiccation, freezing, and radiation, with implications for neuroprotection and plasticity.

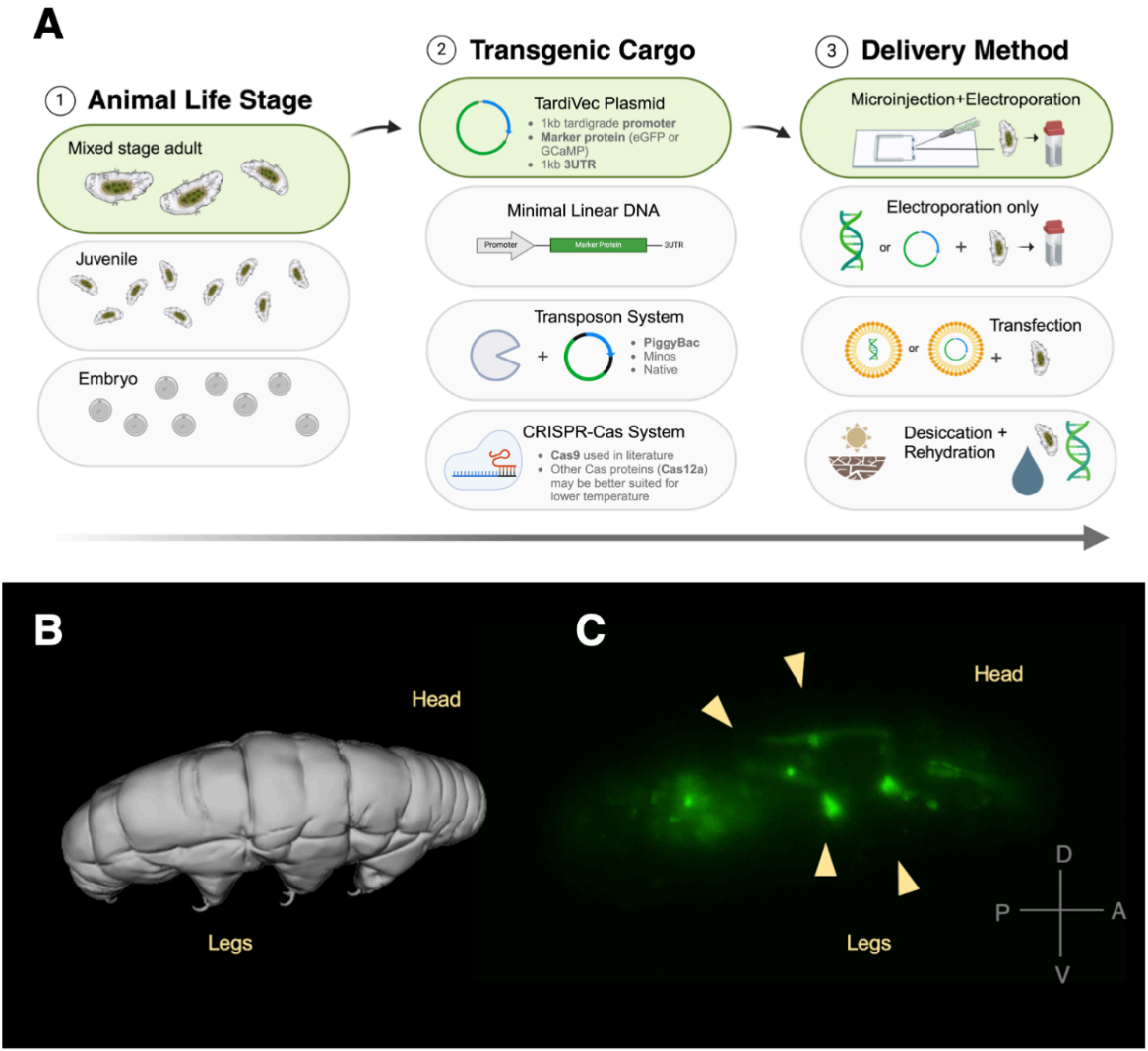

**Figure 4. Key considerations and outcomes for optimizing transgenics in tardigrades**

(A) Flowchart summarizing the major parameters for optimizing a transgenics protocol in tardigrades, including animal life stage (embryo, juvenile, or adult), type of transgenic cargo (e.g., TardiVec plasmid, minimal linear DNA, transposon system, or CRISPR-Cas system), and delivery methods (e.g., microinjection, electroporation, transfection, or desiccation-rehydration).

(B) A 3D rendering of *H. exemplaris* highlighting the orientation of the adult used in transgenics experiments.

(C) Fluorescence image of *H. exemplaris* electroporated with a pHeActin-eGFP TardiVec vector, showing localized eGFP expression in muscle fibers (yellow arrows). Published TardiVec methods (PNAS, Tanaka et al., 2023) do not currently achieve systemic genome editing or germline incorporation. D: dorsal, V: ventral, A: anterior, P: posterior.

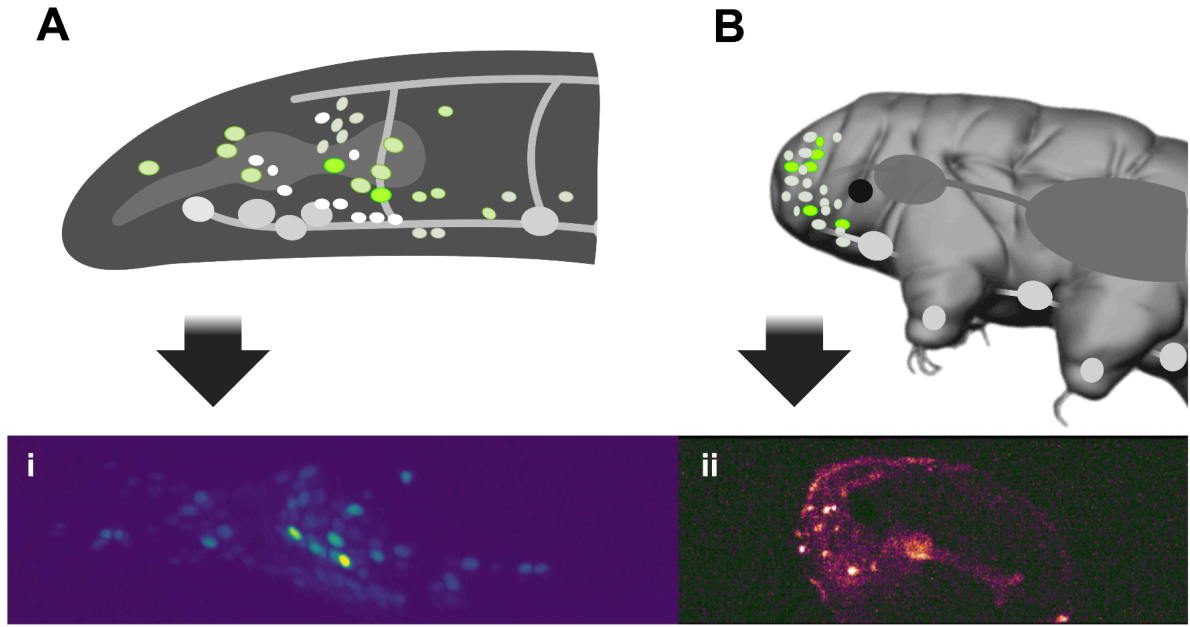

**Figure 5. Adapting live-imaging systems neuroscience methods from *C. elegans* to tardigrades**

(A) Simplified diagram of a *C. elegans* head, illustrating nuclear-localized genetically encoded calcium indicators (GCaMPs) for live neuronal imaging. *i.* Still-frame of nuclear-localized GCaMP6s fluorescence in a live *C. elegans* brain, courtesy of Raymond Dunn, demonstrating established methods for systems neuroscience.

(B) Conceptual illustration of a tardigrade brain and nervous system. Circles represent nuclei of a subset of putative neurons and ganglia. *ii*. Live imaging of an *H. exemplaris* brain using the calcium dye Fluo-4 (10 μM concentration) under fluorescence confocal microscopy at 100× magnification, showcasing the potential for calcium imaging in tardigrades.

**APPENDIX I. In-situ hybridization methods**

Recently, *in situ* hybridization methods using fluorescent probes to target RNA transcripts have emerged as a customizable approach for visualizing putative neuronal markers in non-model organisms, including tardigrades. Technologies like hybridization chain reaction (HCR) and RNAscope have been pivotal in advancing these methods, providing robust and accessible tools for non-model organism research (Choi et al., 2018; Wang et al., 2012). Compared to immunohistochemistry and electron microscopy or other advanced imaging methods, *in situ* hybridization is often easier to implement because it does not require species-specific antibodies or the intricate sample preparation needed for ultrastructural imaging. Instead, it relies on knowledge of the target gene sequence to design specific probes, making it highly adaptable to non-model organisms, as illustrated by unpublished data visualizing the expression of *elav* in a *H. exemplaris* hatchling ([**Figure 2**](#)C). However, these methods have shortcomings, including the inability to directly confirm protein expression or localization and the potential for signal variability due to probe hybridization efficiency or RNA degradation during sample preparation.

Examples of this in situ technology used in tardigrades are recent and include studies in various life stages of *H. exemplaris*. For example, HCR fluorescence in situ hybridization (HCR-FISH) has been applied to juveniles and adults to reveal the expression patterns of genes associated with molting and neuropeptide signaling (Yamakawa and Hejnol, 2024). Specifically, HCR localized molting-related transcripts, such as Sad and Ecr1, to claw glands and neural centers in the head and trunk ganglia, providing insights into the regulation of ecdysis and its neural control. Additionally, this method was used to identify pigment-dispersing factor (PDF) neuropeptide transcripts and their receptor across brain lobes, trunk ganglia, and other tissues, highlighting the multifunctional roles of PDFs in tardigrade neural processing, locomotion, and circadian regulation (Dutta et al., 2024).

Furthermore, *in situ* hybridization using other fluorescent probe-based techniques has revealed spatial patterns of neuronal markers in developing embryos of *H. exemplaris*, allowing researchers to map the organization of the brain and ventral nerve cord during early segmentation and elongation (Smith et al., 2024). These studies demonstrate the adaptability of these tools in developmental and functional investigations of tardigrade neurobiology.